# Fractional parametric resonance in spintronic diodes

Andrea Grimaldi[1,2], Denys Slobodianiuk[3], Eleonora Raimondo[2,4], Raghav Sharma[5,6], Anna Giordano[7], Mario Carpentieri[1], Hyunsoo Yang[5], Riccardo Tomasello[1], Roman Verba[3], Giovanni Finocchio[2,*]

[1] *Department of Electrical and Information Engineering, Politecnico di Bari, Bari, 70126, Italy*
[2] *Department of Mathematical and Computer Sciences, Physical Sciences and Earth Sciences, University of Messina, Messina, 98166, Italy*
[3] *V. G. Baryakhtar Institute of Magnetism of the NAS of Ukraine, Kyiv, 03142, Ukraine*
[4] *Istituto Nazionale di Geofisica e Vulcanologia, Rome, 00143, Italy*
[5] *Department of Electrical and Computer Engineering, National University of Singapore, Singapore, 117583, Singapore*
[6] *Department of Electrical Engineering, Indian Institute of Technology Ropar, Rupnagar, 140001, India*
[7] *Department of Engineering, University of Messina, Messina, 98158, Italy*

[*]corresponding author: gfinocchio@unime.it

## Abstract

Parametric pumping is a powerful tool for the excitation, amplification, and processing of oscillations and waves of different nature. In general, parametric resonance can occur when the pumping frequency $f_p$ and eigenfrequency of a linear mode (or wave) $f_0$ satisfy the relation $f_p \approx 2f_0/n$ ($n = 1, 2, 3, ...$). While such parametric resonance is well known in mechanical, superconductive, and quantum systems, in magnetic and spintronic systems only the lowest ($n = 1$) parametric resonance at double the spin wave mode frequency $f_p \approx 2f_0$ was thoroughly studied and explored. Here, using a theoretical analysis based on both micromagnetic simulations and an analytical model, we show the emergence of resonances at fractional frequencies $f_p \approx 2f_0/n$ (with $n > 10$) in spintronic diodes driven by the simultaneous action of ac spin-transfer torque (STT, current densities < $10^6$ A/cm$^2$) and voltage-controlled magnetic anisotropy (VCMA, effective anisotropy fields < 50 mT). The analytical model shows that parametric magnetization dynamics is irreducible to the standard Mathieu model of a parametric oscillator and demonstrates the crucial role of VCMA-driven mode frequency modulation: together with parametric coupling, it results in higher-order odd ($n = 3, 5, 7, ...$) fractional resonances, observed above certain VCMA pumping threshold, while simultaneous action with linear STT drive produces thresholdless even ($n = 4, 6, 8, ...$) resonances. This higher-order parametric dynamics is not restricted to VCMA pumping and opens new directions for the application of spintronic diodes for nonlinear signal processing and electromagnetic energy harvesting.

## Main text

Spintronic technology provides nanoscale devices with functionalities complementary to CMOS technology [1,2], enabling applications ranging from computing [3–5] to microwave technology [2,6]. Resonant phenomena play a central role in enhancing and controlling the dynamics of such devices, whose behavior is fundamentally governed by the ferromagnetic resonance (FMR) frequency [7–9]. Magnetic tunnel junctions (MTJs), working as spin-torque diodes, biased by a dc current (active regime), exhibit rectification sensitivities overcoming the intrinsic limit of Schottky diodes [1,10,11]. In addition to standard spin-transfer torque (STT), MTJs often exhibit voltage-controlled magnetic anisotropy (VCMA) [12,13]. For passive regimes, VCMA produces a sensitivity well above 1 kV/W [10]. Current efforts focus on the development of compact and more efficient solutions combining arrays of electrically coupled MTJs and co-integration of MTJs with electromagnetic [14] and magnetoelectric antennas [15].

Beyond sensitivity optimization, VCMA is also a key element to electrically drive parametric resonance by an excitation at twice the FMR frequency [16–18]. In magnetic systems, such parametric resonance is well studied and has also been achieved using oscillating magnetic fields [19,20], STT [21], magnetoelastic fields [22], providing different strategies to excite strongly nonuniform and/or large-amplitude oscillations, spin wave amplification, manipulation, and processing.

However, from a general theory of oscillations in dynamical systems, parametric resonance is not restricted to the case with the pumping frequency approximately equal to twice the resonance frequency, $f_p \approx 2f_0$. Within the familiar Mathieu model of parametric oscillator, the parametric instability (leading to parametric resonance) can occur under the condition $f_p \approx 2f_0/n$, where integer number $n = 1,2,3,\ldots$ is the number of so-called Mathieu instability zones [23–25]. Such fractional parametric resonances have been observed in mechanical [26], superconducting [27], quantum systems [28], and others. In magnetic systems, however, magnetization dynamics in higher-order parametric instability zones remains unexplored experimentally and has never been investigated theoretically in detail. It was generally believed that the relatively low Q-factor of magnetic systems, which results in high excitation thresholds in higher-order zones, causes higher-order parametric dynamics to be inaccessible except under specific conditions, like operation close to the instability of the magnetic ground state at low fields [29].

In this work, we theoretically demonstrate that resonances at fractional frequencies are not exotic features, and they can be observed in spintronic diodes owing to the high efficiency of VCMA. Moreover, simultaneous action of microwave STT and VCMA leads to a thresholdless appearance of even higher-order resonances ($n = 4,6,8,\ldots$, i.e. at $f_p \approx f_0/2, f_0/3, f_0/4, \ldots$), which are strongly enhanced under static magnetization tilt from the MTJ normal. Conversely, odd higher-order resonances ($n = 1,3,5,\ldots$, i.e. at $f_p \approx 2f_0, 2f_0/3, 2f_0/5, \ldots$) can only appear after VCMA pumping overcomes a threshold. In addition to these observations, we illustrate the limitation of the Mathieu model for parametric magnetization dynamics and develop a more complete theoretical model.

The prediction of fractional parametric resonance in spintronic devices has important implications for the study of strongly nonlinear dynamics and, from a technological viewpoint, enables the realization of nanoscale, ultralow power and CMOS-compatible devices for nonlinear signal processing (e.g. frequency conversion, subharmonic generation) and electromagnetic energy harvesting.

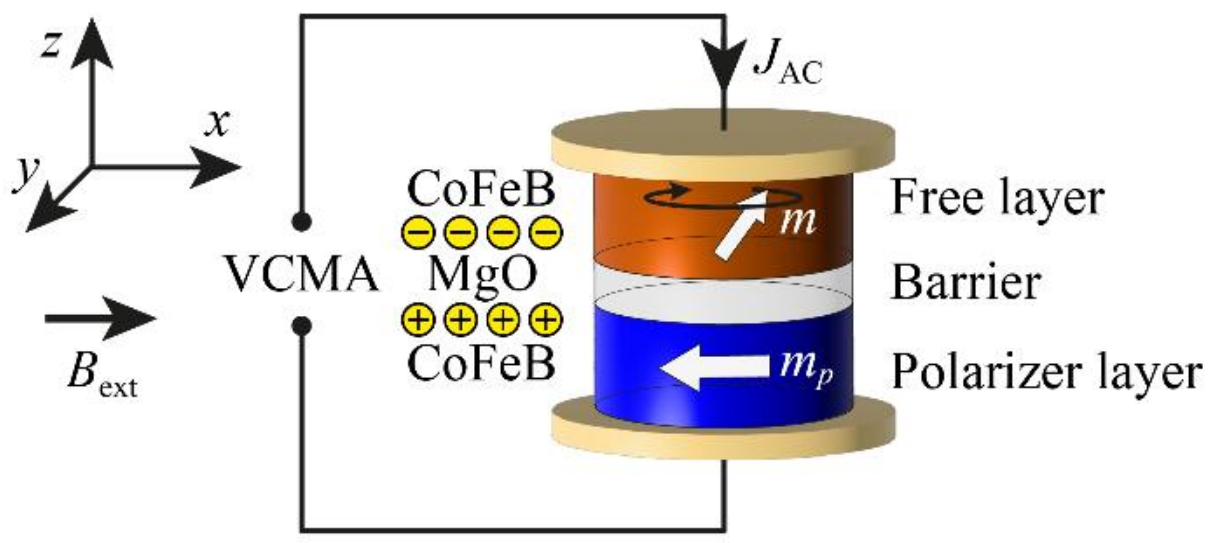

Figure 1. Schematic representation of the simulated device.

We consider a standard MTJ, as shown in Figure 1. The polarizer is aligned to $-\boldsymbol{e}_x$, and the free layer magnetization oscillates out-of-plane around the $\boldsymbol{e}_z$ direction. The MTJ is supplied with ac current $J_{\mathrm{ac}}$ which enables the STT and interfacial VCMA. An in-plane external field $B_{\mathrm{ext}}$ is optionally applied along $\boldsymbol{e}_x$. We perform a systematic study of magnetization dynamics in the MTJ using micromagnetic simulations (see Supplementary Note 1 [30] for the micromagnetic model and parameters).

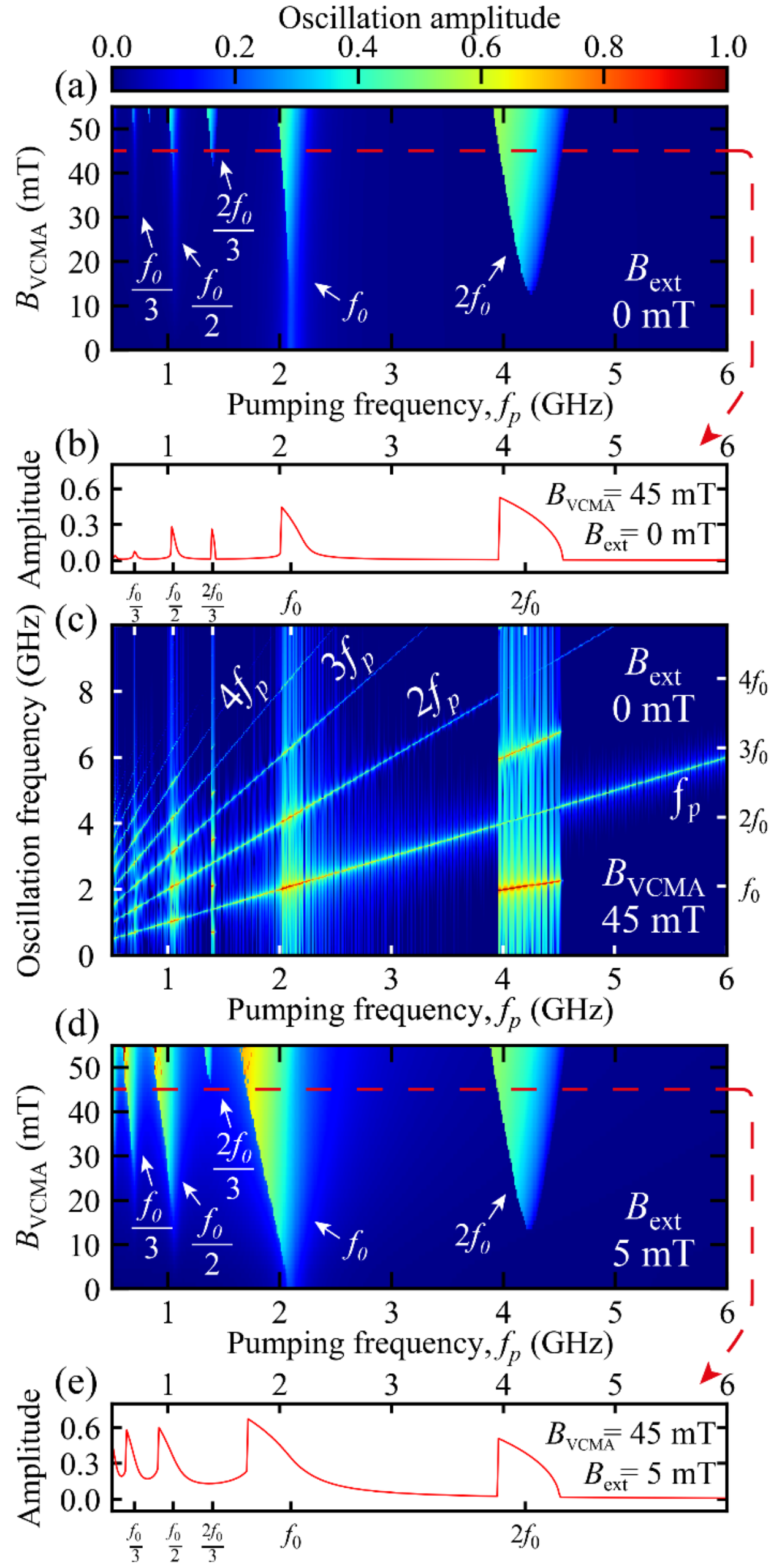

Figure 2. Micromagnetic simulations showing the resonance spectra of the MTJ. (a) Resonance spectra (oscillation magnitude of $m_x$ magnetization component) as a function of $B_{\mathrm{VCMA}}$ and $f_p$ with no external field. (b) The spectrum with $B_{\mathrm{VCMA}} = 45$ mT from (a), the $x$-axis ticks marking the pumping frequency in terms of fractions of the FMR frequency. (c) Fourier spectrum of magnetization dynamics at $B_{\mathrm{VCMA}} = 45$ mT. (d) Resonance spectra as in (a) with $B_{\mathrm{ext}} = 5$ mT. (e) Spectrum at $B_{\mathrm{VCMA}} = 45$ mT from (d).

Figure 2 summarizes the simulated response of MTJ for a fixed $J_{\mathrm{ac}} = 0.1$ MA/cm$^2$ while sweeping the VCMA amplitude $B_{\mathrm{VCMA}}$ from 0 to 55 mT. First, we consider the zero-field case in Figure 2(a). Beyond the FMR ($f_p = f_0$) and conventional parametric resonance at $f_p = 2f_0$, we observe a sequence of additional resonance bands at fractional values of the FMR frequency: $2f_0/3$, $f_0/2$, $f_0/3$. All these fractional bands, as well as the main parametric resonance, are absent when the VCMA contribution vanishes. Only the linear FMR exists at $B_{\mathrm{VCMA}} = 0$, excited by the ac STT. With increasing VCMA amplitude, the bands behave differently: conventional parametric resonance at $2f_0$ and the resonance at $2f_0/3$ exhibit clear threshold behavior, whereas the oscillation amplitude within the resonances $f_0/2$ and $f_0/3$ are thresholdless and evolve much more smoothly with increasing VCMA magnitude.

Additional features can be identified from the resonance spectra at a fixed VCMA, as illustrated in Figure 2(b) for $B_{\text{VCMA}} = 45$ mT (see Supplementary Note 3 [31] for different VCMA magnitudes). Conventional parametric resonance at $2f_0$ exhibits a familiar triangular-shaped curve (the slope direction is given by the negative nonlinear frequency shift in this system) with well-defined edges [18,21], corresponding to the threshold behavior as a function of pumping frequency. The same stands for odd fractional resonance at $2f_0/3$ (through the paper, we classify fractional resonances by the number $n$ like the standard Mathieu oscillator terminology, i.e. $f_p = 2f_0/n$). In contrast, resonance curves at fractional frequencies $f_0/2$ and $f_0/3$ (i.e. resonances corresponding to even zone number $n = 4, 6$) have smooth edges, like the nonlinear FMR curve.

To clarify the underlying physics, we analyze the time-domain traces using fast Fourier transform (FFT) analysis at fixed parametric pumping amplitude. In Figure 2(c), the data from panel (b) is analyzed in the frequency domain. On the $x$-axis, the pumping frequency; on the $y$-axis, the magnetization oscillation frequency, while the color represents the FFT intensity. First, the pumping frequency $f_p$ and its multiples, $2f_p$, $3f_p$, …, are always present in the oscillation spectrum, even if that $f_p$ does not excite a resonant response. This can be inferred from the radial lines that cross Figure 2(c). This is consistent with the fact that microwave STT at $f_p$ linearly excites magnetization oscillations at the same frequency, and harmonics appear due to frequency mixing provided by microwave VCMA (nonlinearity may also contribute to some of the harmonics).

When any of these multiples ($2f_p$, $3f_p$, …,) are equal to the FMR frequency ($qf_p = f_0$), we observe an even zone resonance behavior. In these cases, the peak at FMR frequency $f_0$ is the most prominent, however, peaks at $qf_p \neq f_0$ are also pronounced. This brings us to the conclusion that both the STT-driven linear excitation and frequency mixing by VCMA are relevant in the even zones.

Behavior differs in the odd zones, e.g. the ones at $2f_0/3$ and at $2f_0$ in Figure 2(c). Here, the oscillations at $f_p$ are barely excited, while the most prominent ones are the peak at the FMR frequency and peaks at $f_0 \pm qf_p$ (i.e. $3f_0$ in the first zone, $f_0/3$ and $5f_0/3$ in the third zone, etc.). Therefore, linear STT does not contribute to the odd zone dynamics and only frequency mixing by microwave VCMA is significant.

We also performed simulations under in-plane field $\boldsymbol{B}_{\text{ext}} = B_{\text{ext}}\boldsymbol{e}_x$. Resonance spectra for $B_{\text{ext}}$ = 5 mT are shown in Figure 2(d) (see Supplementary Note 2 [32] for additional $\boldsymbol{B}_{\text{ext}}$ values). When comparing these spectra to the zero-field case, it is found that the static magnetization tilt produced by $\boldsymbol{B}_{\text{ext}}$ does not affect the odd zone resonances (see peaks at $2f_0$ and $2f_0/3$). In contrast, oscillation amplitude in even zones is evidently enhanced, and we can resolve even more fractional resonances, at least up to $f_0/8$. Thus, bias field sensitivity constitutes another prominent difference in odd and even zone resonances. The external field has also the effect of broadening the low-frequency even zones, producing a finite oscillation amplitude across the entire frequency range at large enough VCMA, as observed at $B_{\text{VCMA}} = 45$ mT in Figure 2(e). At larger values of VCMA (not shown), the rectification response exhibits a broadband response at low frequencies similarly to Refs. [14,33].

To explain these results, we develop a theoretical model for the MTJ dynamics. Using spin wave perturbation formalism [34] and Hamiltonian formalism for nonlinear dynamics [35], we derive the following dynamic equation for the dimensionless FMR mode amplitude $c(t)$ (see Supplementary Note 4 [36] for the full derivation):

$$\frac{dc}{dt} + i\left(\omega_0 + T|c|^2 - 2Wb_p(t)\right)c + \Gamma c = 2iVb_p(t)c^* + f_{\text{lin}}e^{-i\omega_p t}. \tag{1}$$

Here $\omega_0 = 2\pi f_0$ is the FMR angular frequency, $T$ is the nonlinear frequency shift and $\Gamma$ is the damping rate. The linear force, acting on the magnetization at the frequency $\omega_p = 2\pi f_p$, comes from both the STT and VCMA,

$$f_{\text{lin}}(t) = \frac{\sigma_{\text{STT}} g_T J_{\text{ac}} \sqrt{\epsilon}}{2\sqrt{2}} \cos\theta + i\frac{\gamma B_{\text{VCMA}}}{4\sqrt{2\epsilon}} \sin 2\theta\,, \tag{2}$$

where $\theta$ is the static magnetization tilt angle, $\epsilon$ is the ratio of dynamic magnetization components describing the magnetization precession ellipticity, $g_T$ is the Slonczewski polarization function, $\sigma_{\text{STT}}$ is the STT efficiency, and $\gamma$ is the modulus of gyromagnetic ratio. In addition, modulation of the anisotropy via VCMA results in the appearance of two parametric-like terms, $2iVb_p(t)c^*$ and $2iWb_p(t)c$, which, hereafter, we refer to parametric coupling and the frequency modulation, respectively. Here, the pumping field $b_p(t) = B_{\text{VCMA}}\cos(\omega_p t)$ is the VCMA modulation of effective field, and the parametric and frequency modulation coefficients are given by

$$V = \frac{\gamma}{4\epsilon}\left((1-\epsilon^2)\cos^2\theta - \sin^2\theta\right), \tag{3}$$

$$W = \frac{\gamma}{4\epsilon}\left((1+\epsilon^2)\cos^2\theta - \sin^2\theta\right). \tag{4}$$

Equation (1) differs from the standard model of parametric resonance of magnetization [17,18,20] by the frequency modulation term. This term has no effect on the parametric dynamics in the first instability zone and has therefore been omitted in previous works. However, for fractional resonances it is of a principal importance, as shown below. Notably, Eq. (1) is irreducible to the forced Mathieu parametric oscillator model. The principal difference is in the number of parametric-like terms: in the Mathieu model, only the frequency modulation is present ($\ddot{c} + \omega_0^2[1 + \xi\cos(\omega_p t)] + \cdots$), which is sufficient for describing the parametric dynamics in all instability zones; for the magnetization dynamics, the pumping produces two distinct effects.

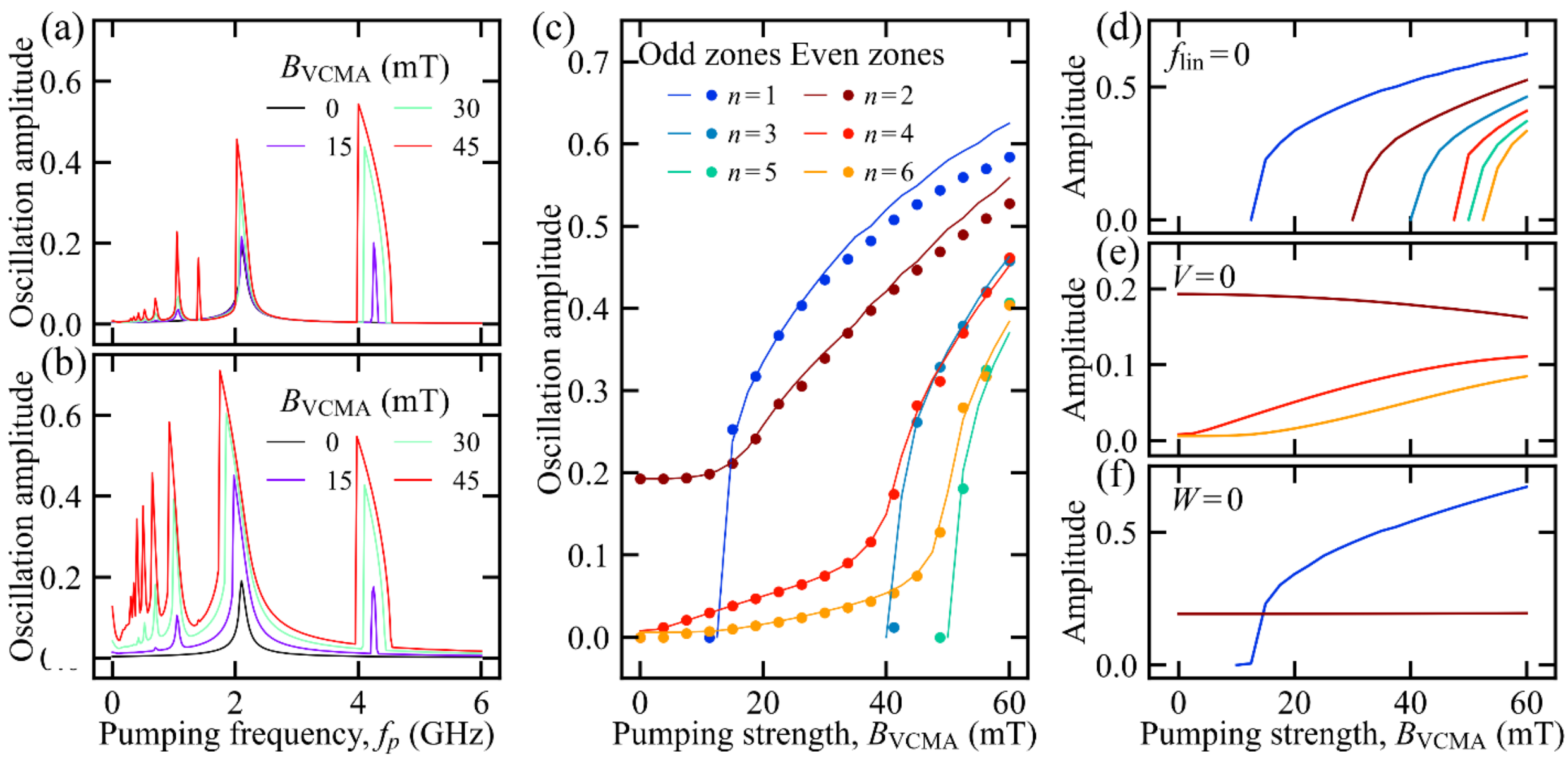

Figure 3. Results of theoretical model. (a,b) Resonance spectra of MTJ for different VCMA pumping strength at external field $B_{\rm ext} = 0$ (a) and $B_{\rm ext} = 5$ mT (b). (c) Amplitudes of resonance peaks in different resonance zones for $B_{\rm ext} = 0$ highlighting different behavior of odd and even zone resonances; lines – solution of Eq. (1), symbols – micromagnetic simulations. (d,e,f) Stationary solution of Eq. (1) for no linear force, $f_{\rm lin} = 0$, (d), no parametric coupling term, $V = 0$ (e), and no frequency modulation term, $W = 0$ (f), legend is shared with (c) for all three panels.

We solve Eq. (1) numerically and plot resonance spectra (stationary solutions) under different pumping strengths and bias field intensities, as shown in Figure 3 (a) and (b). These spectra possess all the principal features observed in the micromagnetic simulations: appearance of fractional resonances, threshold and thresholdless behavior of odd and even fractional resonances, respectively, and their different sensitivity to external field.

From the micromagnetic resonance spectra of Figure 2(a), we extract the peak amplitude dependence on VCMA in different bands, see the markers in Figure 3(c). When estimating the same quantity using the theoretical model (full lines in the figure), we observe that there is a remarkable correspondence with the micromagnetic simulations. Again, while odd resonances possess a threshold that, similarly to the Mathieu model, increases with the resonance number $n$, even resonances are thresholdless. Additionally, at high VCMA strength we observe a bend of the $m_x(B_{\rm VCMA})$ dependence and an increase of the oscillation amplitude, indicating switching to a different regime, as clarified below.

The model also provides insight into the magnetization dynamics. By switching off the linear force, $f_{\rm lin} \rightarrow 0$, see Figure 3(d), we observe no change in behavior for odd zone resonances. In contrast, even zone resonances change drastically, becoming like the odd ones, and appear only when a certain threshold value of the VCMA pumping is overcome. Thus, in the absence of linear force we get a qualitatively identical picture to the Mathieu model, a set of higher-order parametric instability zones. Overcoming the parametric instability threshold explains the bends of oscillation amplitude vs pumping field dependence, when both linear force and parametric pumping are applied (see Figure 3(c)).

Further changes in the resonance behavior occur if we set the parametric coupling to zero, $V \rightarrow 0$, see Figure 3(e), retaining only frequency modulation and linear force. We still observe thresholdless excitation in even zones, while odd zone resonances vanish completely. When linear force is absent, no magnetization dynamics is observed.

In the opposite case, if frequency modulation is neglected ($W \to 0, V \neq 0, f_{\text{lin}} \neq 0$), only linear FMR and conventional parametric resonance at the double FMR frequency are observed, as shown in Figure 3(f).

In summary, nonzero quantity $V$ is a pre-requisite for the observation of spontaneous (without linear force contribution) parametric resonance in any parametric instability zone. Thus, we refer to $V$ as the parametric coupling coefficient. Spontaneous resonance in any higher-order zone, $n \geq 2$, requires the presence of both the parametric coupling and frequency modulation. However, forced parametric resonance in even fractional resonance zones is possible even in the absence of parametric coupling, with only frequency modulation and linear force. In other words, forced fractional parametric resonance can be observed in magnetic systems where spontaneous parametric resonance is prohibited. This constitutes a principal difference with forced Mathieu oscillator model [25].

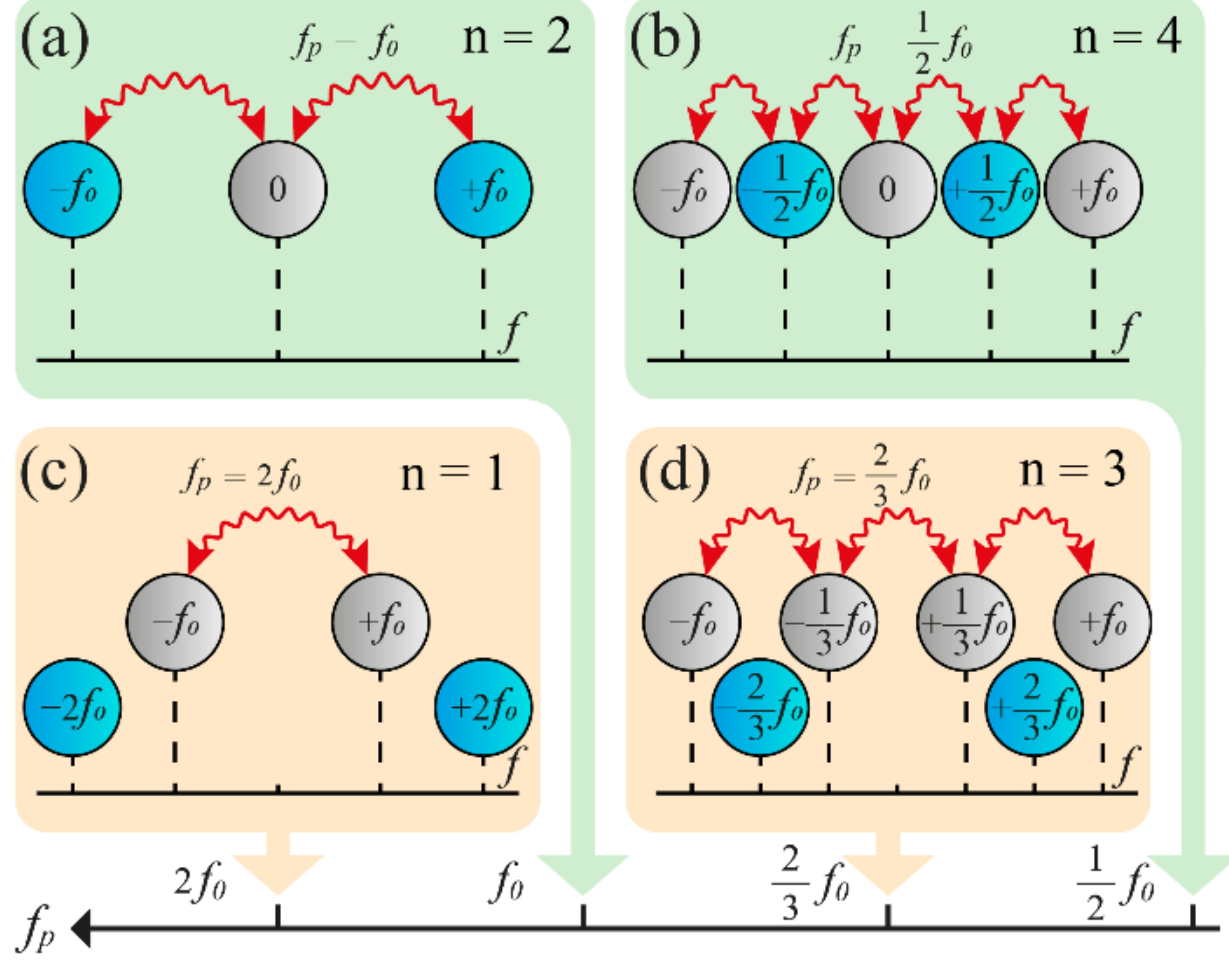

Figure 4. Scheme of pumping coupling interactions in the first four stable zones. (a,b) In even zones, the pumping frequency modes are included in the interaction chain coupling the FMR modes. (c,d) In odd zones, the interaction chain that couples the FMR modes does not include the pumping frequency mode.

A schematic illustration of spontaneous and forced parametric dynamics is presented in Figure 4. Spontaneous parametric instability requires harmonics at $f_0$ and $-f_0$ to be coupled. In the $n$-th zone, it is achieved via $n$ interactions with the pumping, each of which produces harmonics shifted by $\pm f_p$ compared to the initial one. In even zones, harmonic at $f_p$ is coupled to the resonant $f_0$ harmonic via $(n/2 - 1)$ interactions, and we observe resonantly amplified behavior even far from higher-order instability threshold, as shown in Figure 4(a) and (b), for $n = 2$ and 4. In contrast, in odd zones, the pumping frequency is excluded from the set of interacting harmonics and no interaction with the pumping can transfer weak excitation at $f_p$ by linear force to the resonance FMR harmonic, as shown in Figure 4(c) and (d), for $n = 1$ and 3.

Importantly, the VCMA fields and ac current densities considered here are within experimentally achievable ranges [33,37]. Special highlight should be given to even zone resonances under combined linear force and parametric pumping. First, as VCMA produces linear force in a tilted static magnetic state (see Eq. (2)), fractional resonances at $f_0/q$ could be observed even in a single ferromagnetic layer-dielectric structure exhibiting VCMA. This additional linear contribution (exceeding the STT ones by 8 times at $B_{\text{ext}} = 5$ mT, see Supplementary Note 4 [36]) is the reason of strong fractional resonance amplification under an applied bias field (see Figure 2(d) and Figure 3(b)). Another important feature is that forced fractional resonance requires only $(q - 1)$ frequency conversions (compared to $n = 2q$ conversions for spontaneous resonance), all of which are achieved via the frequency modulation term. In contrast to parametric coupling, which requires specific symmetry of magnetization oscillations and pumping (e.g. elliptic precession, $\epsilon \neq 1$, for perpendicular magnetization, see Eq. (3)), the frequency modulation coefficient is almost always nonzero, and it is much larger (>4 times in our case). While ellipticity vanishes when FMR frequency increases, making observation of spontaneous parametric resonance challenging (with microwave magnetic field pumping; VCMA parametric coupling in a tilted state is not subject to such restrictions), frequency modulation remains large. Therefore, observation and utilization of forced fractional parametric resonances is expected to be relatively simple in a wide range of magnetic nanosystems.

In summary, we have shown the appearance of fractional resonances at $2f_0/n$ frequencies in a spintronic MTJ under simultaneous action of microwave STT and VCMA. Odd-$n$ resonances are spontaneous higher-order parametric resonances; they are observed above their respective $n$-dependent thresholds of VCMA magnitude, which lie in the range of tens of mT effective VCMA field, accessible in state-of-the-art MTJs. Even fractional resonances arise from the interplay between STT linear force and frequency modulation induced by VCMA, which leads to their thresholdless appearance. The VCMA contribution to the linear force in a tilted static state greatly enhances even fractional resonances and eventually leads to a broadband MTJ response. The developed theoretical model predicts that such fractional resonance behavior is not a unique feature of MTJs under combined action of STT and VCMA, and should also occur in other magnetic nanosystems, provided a simultaneous action of frequency modulation with parametric coupling (for spontaneous resonances) or with linear force (for even-zone forced resonances). Our results establish a general framework for engineering fractional resonances in spintronic devices and open new routes for frequency multiplication, frequency-selective microwave detection and signal processing, and broadband energy harvesting.

**Acknowledgements**

This work has been supported by Project No. 101070287, SWAN-on-chip, HORIZON-CL4-2021-DIGITALEMERGING-01 and by the PETASPIN association (www.petaspin.com). R.V. acknowledges support by National Research Foundation of Ukraine, Grant No. 2025.03/0005.